\documentstyle[11pt,epsf]{article}  
\begin{document}  
  
\newcommand\beq{\begin{equation}}  
\newcommand\eeq{\end{equation}}  
\newcommand\bea{\begin{eqnarray}}  
\newcommand\eea{\end{eqnarray}}  
  
\markboth{Eckhardt}{Eigenvalue statistics in quantum ideal gases}

\title{EIGENVALUE STATISTICS IN QUANTUM IDEAL GASES}  
\author{BRUNO ECKHARDT  
\thanks{Fachbereich Physik, 
Philipps Universit\"at Marburg, 35032 Marburg, Germany}
}  
\maketitle     
               
\begin{abstract}                     
The eigenvalue statistics of quantum ideal gases  
with single particle energies $e_n=n^\alpha$ are studied. A recursion  
relation for the partition function allows to calculate the mean  
density of states from the asymptotic expansion for the single particle   
density. For integer $\alpha>1$ one expects and finds number theoretic  
degeneracies and deviations from the Poissonian spacing distribution.  
By semiclassical arguments, the length spectrum of the classical system  
is shown to be related to sums of integers to the power $\alpha/(\alpha-1)$.  
In particular, for $\alpha=3/2$, the periodic orbits are related  
to sums of cubes, for which one again expects number theoretic  
degeneracies, with consequences for the two point correlation function.  
\end{abstract}   
   
%\begin{keywords}   
%Quantum ideal gases, level spacing distributions, form factor  
%\end{keywords}  
  
\section{Introduction}  
Most investigations of quantum chaos have focussed on the   
effects in single particle systems. The prime examples of  
frequently studied systems, such as hydrogen in a magnetic   
field, the standard map or small molecules all belong to this  
class (Eckhardt 1988, Casati and Chirikov 1995). 
Even electrons in a solid, a standard many body system,  
has until recently been reduced to a single (quasi)particle   
system. Yet the study of many-body quantum systems, even if they  
are integrable  can be interesting for several reasons.  
  
For once, the spin-statistics theorem, which requires   
quantum wave functions to be either totally symmetric   
(Bose-Einstein statistics) or totally  
antisymmetric (Fermi-Dirac statistics) under exchange of  
particles changes the spectrum compared to the simple   
Maxwell-Boltzmann type superposition of the individual   
single particle density. This gives rise to changes in the   
total density of states, as is well known in statistical  
mechanics. The full program of implementing the permutation  
symmetry semiclassically in the mean density of states and   
in the trace formulas has recently been taken up  
by Weidenm\"uller et al. (1993a,b).  
  
Experimentally, small clusters are examples of systems with  
many degrees of freedom for which one should use microcanonical  
averaging rather than canonical, since the number of atoms  
or electrons is rather fixed. The difference between   
microcanonical and canonical ensembles can be observed.  
The spectral statistics of small systems influences their  
thermodynamical behavior (M\"uhlschlegel 1991).  
  
Furthermore, symmetric systems can serve as a reference point for 
systems with weakly broken symmetries. In particular, it has been 
observed that weak symmetries, perhaps of dynamical origin, 
can give rise to strong degeneracies in spectra, the 
Shnirelman peak (Shnirelman 1993, Chirikov and Shepelyansky 1995). 
 
In this contribution only integrable ideal gases will  
be analyzed. The energy levels are constrained to be   
a power of the quantum number,  $E_n=n^\alpha$, where   
$n=1,2,3,\ldots$ and $\alpha>1$, except where noted.   
This family includes for instance the eigenenergies for a   
particle of mass $m$ confined to a 1-d box  
of width $L$, measured in units of $4\pi^2\hbar^2/2mL^2$. It also describes  
the asymptotic eigenvalues for particles in homogeneous potentials of   
degree $p$, for which $\alpha=2p/(p+1)$ (Seligman et al. 1985, 
Seligman and Verbaarschot 1987). The harmonic case, $\alpha=1$,  
causes problems for the stationary phase approximations used below and  
will only be considered occasionally. The deviations of its level  
spacing distribution from Poissonian have been studied at length previously 
(Berry and Tabor 1977a). 
The quantum ideal gases then have   
the eigenvalues  
\beq  
E = \sum_{n_1,\ldots,n_D} n_i^\alpha \mbox{\quad with\ } \left\{  
{\matrix{ \mbox{no constraint} & \mbox{Maxwell-Boltzmann (MB)}\cr  
n_1\le n_2\le \ldots \le n_D & \mbox{Bose-Einstein (BE)} \cr  
n_1 <  n_2 <  \ldots  <  n_D & \mbox{Fermi-Dirac (FD)} } } \right. \,,  
\eeq  
where the different statistics have been indicated.  
 
The questions addressed here concern the mean density of states 
(section 2), the level spacing distribution in particular   
for integer $\alpha$ (section 3) and the behaviour of the 
pair correlation function (section 4). The results on the 
mean density of states are of a more general nature,  
whereas some examples are particular to the powers 
$\alpha$ considered. The final section contains some 
speculations on the relevance of number theory especially 
for the pair correlation function for some  
rational values of $\alpha$. 
 
\section{The partition function and the mean density of states}  
\subsection{The connection}  
The partition function for a quantum mechanical system is defined  
as   
\beq  
Z(\beta) = \mbox{tr\,}e^{-\beta H} = \sum_j e^{-\beta E_j}\,,  
\eeq  
where the last sum extends over all eigenvalues $E_j$. The density  
of states $\rho(E)$ is related to $Z$ by a Laplace transform,  
so that the poles in an asymptotic expansion of $Z$ for small  
$\beta$ are related to the rate of divergence of $\rho(E)$  
for large energy and thus to te mean density of states.  
In particular, if  
\beq  
Z(\beta) \sim \sum_j c_j \beta^{-\gamma_j}   
\eeq  
then  
\beq  
\overline{\rho}(E)  
\sim \sum_j {c_j \over \Gamma(\gamma_j)} E^{\gamma_j-1} \,.  
\eeq  
As an example, in the harmonic case $\alpha=1$, one has  
\beq  
Z(\beta) = {1 \over e^\beta-1} \sim {1\over \beta} - {1\over 2}  
+ {\beta \over 12} - {\beta^3\over 720} +  \ldots  
\label{ho_exp}  
\eeq  
and for the mean density of states  
\beq  
\overline{\rho}(E) = 1\,.  
\eeq  
Thus, whereas the partition function contains contributions  
from positive powers of $\beta$ (which contain information on  
the shortest periodic orbits in the system, Berry and Howls 1994),  
these terms are   
cancelled in the density of states by the poles of the   
$\Gamma$-function in the denominator.  
  
\subsection{The recursion relation for $D$-particle partition functions}  
For the Maxwell-Boltzmann case with no restrictions on the integer sums,   
the partition function for $D$ particles can be written down explicitly,  
\beq  
Z^{(MB)}_D(\beta) = \left(Z_1(\beta)\right)^D \,.  
\eeq  
This is no longer possible for the symmetry reduced subspaces.  
However, there is a simple recursion relation involving the  
partition functions for all particle numbers up to $D$,  
\beq  
Z_D(\beta) = {1\over D}\sum_{k=1}^D  
(\pm1)^{k+1} Z_1(k\beta) Z_{D-k}(\beta)\,;  
\label{rec_rel}  
\eeq  
the $+1$ applies in the Bose-Einstein subspace and   
the $-1$ in the Fermi-Dirac subspace.  
A combinatorial proof of this relation was given by   
Bormann and Franke (1993). A more direct analytical proof may be  
based on the grand canonical formalism (Reif 1965).  
  
The grand canonical partition function is defined as  
\bea  
\Omega(z,\beta) &=& \sum_{D=0}^\infty z^D Z_D(\beta)\\  
&=& \sum_{D=0}^\infty \sum_{n_1,\ldots,n_D} z^D  
e^{-\beta\sum_i n_i^\alpha}\,;  
\eea  
the last sum on quantum numbers is restricted by the selection rules  
for the different statistics. Passing to an occupation number representation,  
where $g_n$ denotes the number of particles in quantum state $n$, one  
obtains  
\beq  
\Omega(z,\beta) = \sum_{g_1,g_2\ldots} z^{\sum_n g_n}   
e^{-\beta \sum_n g_n n^\alpha} \,.  
\eeq  
In the case of the Bose-Einstein statistics, the occupation numbers can take  
on all nonnegative integer values, and the summation on $g_n$ gives rise   
to geometric series. In the case of the Fermi-Dirac statistics, the   
occupation numbers can only take on the values 0 and 1. The result for   
both cases can be combined in a single expression,  
\beq  
\Omega(z,\beta) = -\prod_{n=1}^\infty \left(  
1-z\theta e^{-\beta n^\alpha}\right)^{-\theta}\,,  
\eeq  
where $\theta=+1$ for Bose-Einstein and $\theta=-1$ for Fermi-Dirac   
statistics.  
  
The $D$-particle partition function follows from taking an $n$-th derivative  
of $\Omega$ with respect to $z$ at $z=0$. The first derivative becomes  
\bea  
{\partial \Omega(z,\beta) \over \partial z} &=&  
\sum_{n=1}^\infty {e^{-\beta n^\alpha} \over 1-\theta z e^{-\beta n^\alpha}}  
\, \, \Omega(z,\beta)\\  
&=& S(z,\beta) \Omega(z,\beta)\,.  
\eea  
The recursion relation (\ref{rec_rel}) now follows from the formula for   
derivatives of a product and the observation that the $k$-th derivative of   
$S$ at $z=0$ is related to the single particle partition function,  
\beq  
\left.{\partial^k S(z,\beta) \over \partial z^k}\right|_{z=0} = \theta^k   
Z_1((k+1)\beta)\,.  
\eeq  
This result holds for other forms of the single particle energies as well.  
  
\subsection{Examples}  
The first few partition functions when reduced to the single  
particle partition function become  
\bea  
Z_2(\beta) &=& {1\over 2} \left(Z_1^2(\beta) \pm Z_1(2\beta)\right)\cr  
Z_3(\beta) &=& {1\over 6} \left(Z_1^3(\beta) \pm 3 Z_1(\beta) Z_1(2\beta)  
+ 2Z_1(3\beta)\right) \\ 
Z_4(\beta) &=& {1\over 24} \left(Z_1^4(\beta) \pm 6 Z_1^2(\beta) Z_1(2\beta)  
+ 8 Z_1(\beta) Z_1(3\beta) \right. \cr   
&\ & \left. + 3 Z_1^2(2\beta) \pm 6 Z_1(4\beta)\right)\nonumber\cr  
&\mbox{etc.}& \nonumber  
\eea  
where the plus signs apply for the BE-statistics and the   
minus signs for the FD-statistics. The corresponding densities of states  
become  
\bea  
\overline{\rho}_2(E)  
&=& {1\over 2} \sum{n_1,n_2} \delta(E-n_1^\alpha-n_2^\alpha)  
\pm {1\over 2} \sum_n \delta (E-2 n^\alpha) \nonumber \\  
\overline{\rho}_3(E)  
&=& {1\over 6} \sum_{n_1,n_2,n_3} \delta(E-n_1^\alpha-n_2^\alpha  
-n_3^\alpha)\\  
&\ & \pm {1\over 2} \sum_{n_1,n_2} \delta(E-n_1^\alpha-2n_2^\alpha)  
+{1\over 3} \sum_n \delta (E-3n^\alpha) \nonumber \\  
&\mbox{etc.}& \nonumber  
\label{dos}  
\eea  
The leading order contribution is $Z^{(MB)}/D!$, with the familiar  
permutation factor $D!$. Thus the mean density of states is to   
leading order the same for Bose-Einstein and Fermi-Dirac statistics  
and equal to $1/D!$ the density of states for the Maxwell-Boltzmann   
statistics. The next to leading order terms correct for the counting  
of states with two or more identical quantum numbers.  
  
Further progress can be made if either the full single particle  
partition function or at least its asymptotic expansion are known.  
Such is the case for $\alpha=1$, the harmonic oscillator,  
and for $\alpha=2$ and $n=0,\pm1, \pm2, \pm3, \ldots$, i.e.  
a particle on a ring.   
  
The single particle partition function for the harmonic energies  
$\alpha=1$ and its asymptotic expansion have been given above  
(\ref{ho_exp}).  
The partition functions for two particles then become  
\bea  
Z^{BE}_2(\beta) &=& {1\over2\beta^2} - {1\over 4 \beta} - {1\over 24}   
+{7\over 1440} \beta^2 \pm \ldots \\   
Z^{FD}_2(\beta) &=& {1\over2\beta^2} - {3\over 4 \beta} + {11\over 24}   
+{7\over 1440} \beta^2 \pm \ldots \,.  
\eea  
For the mean density of states, these expansions imply  
\bea  
\overline{\rho}^{BE}_2(\beta) &=& {1\over2} E - {1\over 4} \\  
\overline{\rho}^{FD}_2(\beta) &=& {1\over2} E - {3\over 4} \,.  
\eea  
Here the correction terms have the same sign but different magnitude. 
 
As a second example we take a particle on a ring (periodic boundary  
conditions on the interval), for which  
\beq  
Z_1(\beta) = \sum_{n=-\infty}^{+\infty} e^{-\beta n^2}   
\sim \left({\pi\over\beta}\right)^{1/2} \,.  
\eeq  
From this one can derive that  
\bea  
Z_2(\beta) &=&{1\over 2}   \left({\pi\over\beta}\right)^{-1} \pm   
           {\sqrt{2}\over4}\left({\pi\over\beta}\right)^{-1/2} \nonumber\\  
Z_3(\beta) &=& {1\over 6}  \left({\pi\over\beta}\right)^{-3/2} \pm   
           {\sqrt{2}\over4}\left({\pi\over\beta}\right)^{-1} +  
           {\sqrt{3}\over9}\left({\pi\over\beta}\right)^{-1/2}\\   
Z_4(\beta) &=& {1\over 24} \left({\pi\over\beta}\right)^{-2} \pm   
           {\sqrt{2}\over8}\left({\pi\over\beta}\right)^{-3/2} \cr  
           &\ & +  
           \left({\sqrt{3}\over9}+{1\over 16}\right)  
                           \left({\pi\over\beta}\right)^{-1} \pm   
           {1\over8}       \left({\pi\over\beta}\right)^{-1/2}\nonumber 
        %   \cr &\vdots&  
\eea  
and the densities  
\bea  
\overline{\rho}_2(E) &=& {\pi\over2} \pm {\sqrt{2}\over 4} E^{-1/2} \cr 
\overline{\rho}_3(E) &=& {\pi\over 3} E^{1/2} \pm   
           {\sqrt{2}\over4}\pi +  
           {\sqrt{3}\over9} E^{-1/2}\\   
\overline{\rho}_4(E) &=& {\pi^2\over 24} E \pm   
           {\sqrt{2}\pi\over4} E^{1/2} +  
           \left({\sqrt{3}\over9}+{1\over 16}\right)\pi \pm   
           {1\over8}       E^{-1/2}\nonumber 
     %   \cr &\vdots& \nonumber  
\eea  
Again, the plus sign refers to the Bose-Einstein statistics and the   
minus sign to the Fermi-Dirac statistics. Evidently, since the   
one particle partition function consists of the the leading order  
term only, all lower order terms are due to symmetrization and the    
signs clearly reflect the statistics. 
  
\begin{figure}[t] 
\epsfverbosetrue 
\epsfxsize=\hsize 
\epsfbox{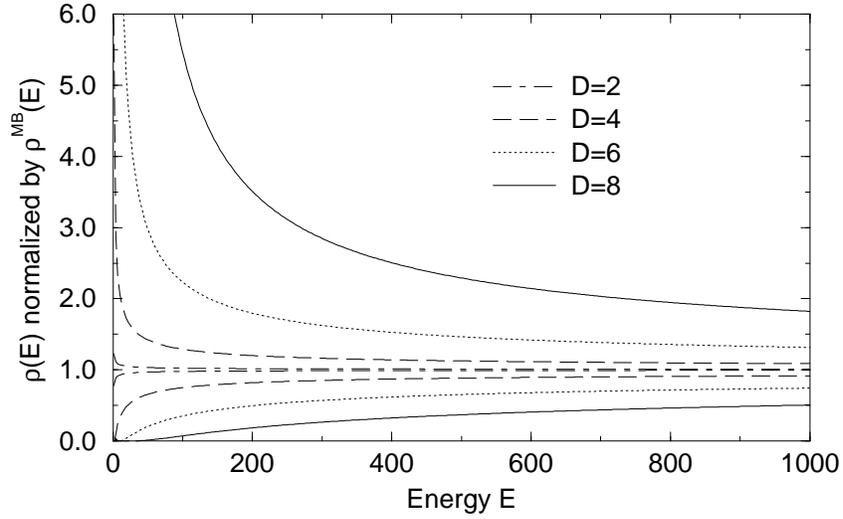}  
\caption[]{  
Mean density of states for $D$-particle systems in the   
Bose (top) and Fermi (bottom) subspaces for particles on  
a ring. The densities are normalized by the desymmetrized  
Maxwell-Boltzmann densities.   
\label{density_fig}}  
\end{figure}

The size of the next to leading order corrections increase  
rapidly with the particle number $D$ and become very important for  
large $D$. As can easily been shown by induction,  
\beq  
Z_D(\beta) = {\pi^{D/2} \over D!} \beta^{-(D/2)}  
\left(1\pm  
c_D \beta^{1/2} \pm \cdots\right)  
\eeq  
so that the mean density of states becomes  
\beq  
\overline{\rho}_D(\beta) = {\pi^{D/2} \over D!\Gamma(D/2)} E^{(D-2)/2}  
\left(1\pm c_D E^{-1/2} \pm \cdots\right) \,.  
\eeq  
with  
\beq  
c_D = {\sqrt{2} D (D-1) \Gamma(D/2) \over 4 \sqrt{\pi}  
\Gamma((D-1)/2)} \sim {1\over 4\sqrt{\pi}} (D-1)^{5/2} \,.  
\eeq  
The energy $E_c$, where $c_D E_c^{-1/2} \sim 1$ thus increases like  
$(D-1)^5$. This may be compared to the groundstate energy  
of the Fermi system, $E_F \sim D^3$. Even when compared to this  
the importance of this term increases like $E_c/E_F\sim (D-1)^{5/3}$.  
The approach to the density of states of the classical ideal gas 
is thus very slow, and it will be difficult to estimate ground  
states accurately from the leading order terms (something that 
works surprisingly well in many cases in few degree of freedom 
systems). Some examples for the behaviour of the next to leading 
order corrections are shown in Fig.~\ref{density_fig}. Similar  
behavior can be found in the harmonic oscillator case.   
  
\subsection{Asymptotic expansions of single particle partition functions}  
For integer $\alpha$ one can use the Euler-MacLaurin summation formula  
to derive an asymptotic expansion for the single particle partition   
function starting from the representation  
\beq  
\sum_{n=1}^\infty e^{-\beta n^\alpha} =  
\int_0^\infty dx\, e^{-\beta x^\alpha} - {1\over 2}  
+ \left. \sum_{k=1}^\infty {B_{2k}\over (2k)!}   
{\partial^{2k-1}\over \partial n^{2k-1}}  
e^{-\beta n^\alpha} \right|_{n=0}\, ,  
\eeq  
where $B_{2k}$ are the Bernoulli numbers, $B_{2k} \sim (-1)^{k-1}  
(2k)! / (\pi^{2k} 2^{2k-1})$.  
The leading order divergence comes from the integral, whereas  
the remaining terms give a power series in $\beta$,  
\beq  
Z_1(\beta) \sim \Gamma\left({\alpha+1\over\alpha}\right) \beta^{-1/\alpha}  
+\sum_j c_j \beta^{-\gamma_j} \,.  
\eeq  
A power series expansion of $\exp(-\beta n^\alpha)$ shows that only 
the powers $\gamma=(2k-1)\alpha$ carry non-zero weights. As noted before,  
these do not contribute to the density of states for the single particle  
but can be brought to live in the many particle system 
through combinations with sufficiently many powers of the  
first term.  
  
For rational $\alpha$ this does not work, since the derivatives  
required in the Euler-MacLaurin formula do not terminate,  
giving diverging coefficients for the powers of $\beta$.   
Numerical results are consistent with the leading order behavior  
being given by the integral and the further terms in the   
desymmetrized version being a power series in $E^{1/\alpha}$.  
  
For later reference, I note the mean density of states for  
$D$ particles that results from this asymptotic expansion,  
\beq  
\overline{\rho}_D(E) = {\Gamma\left({\alpha+1\over\alpha}\right)^D  
\over D! \Gamma\left({D\over\alpha}\right)} E^{(D-\alpha)/\alpha}   
+ \cdots \,.  
\label{density}  
\eeq  
The mean density of states thus decreases for $D<\alpha$, is asymptotically   
constant for $D=\alpha$ and increases for $D>\alpha$.  
  
\section{Nearest neighbor spacings}  
\subsection{Numerical results}  
The Maxwell-Boltzmann gas with its huge average degeneracies because  
of permutation symmetry gives rise to a rather singular level  
spacing distribution,  
\beq  
P(s) = (1-g) \delta(s) + {1\over g^2} e^{-gs}  
\eeq  
where the mean degeneracy $g=D!$. These massive degeneracies will  
obviously not survive perturbations. However, as Shnirelmans observations  
show, the low lying spectra of a weakly perturbed system may still  
have a $\delta$-function contribution because of (almost)  
degeneracies between the perturbed symmetry related tori 
(shnirelman 1993, Casati and Shepelyansky 1995). 
  
Within the symmetry reduced Bose-Einstein and Fermi-Dirac subspaces  
one does not expect this effect and generically finds a Poissonian  
spacing distribution (in accord with Berry and Tabor 1997a).  
However, for integer powers of $\alpha$, all  
the energy levels will be integer and there can be some number theoretic  
degeneracies 
(Casati et al 1985). In the case of $\alpha=2$ it is known that the level  
spacing distribution collapses to a delta function. Thus although  
the density of states is constant, some large integers can be  
expressed in an increasing number of ways as sums of squares, and the   
gaps between them increase logarithmically.  
  
\begin{figure}[t] 
\epsfverbosetrue 
\epsfxsize=\hsize 
\epsfbox{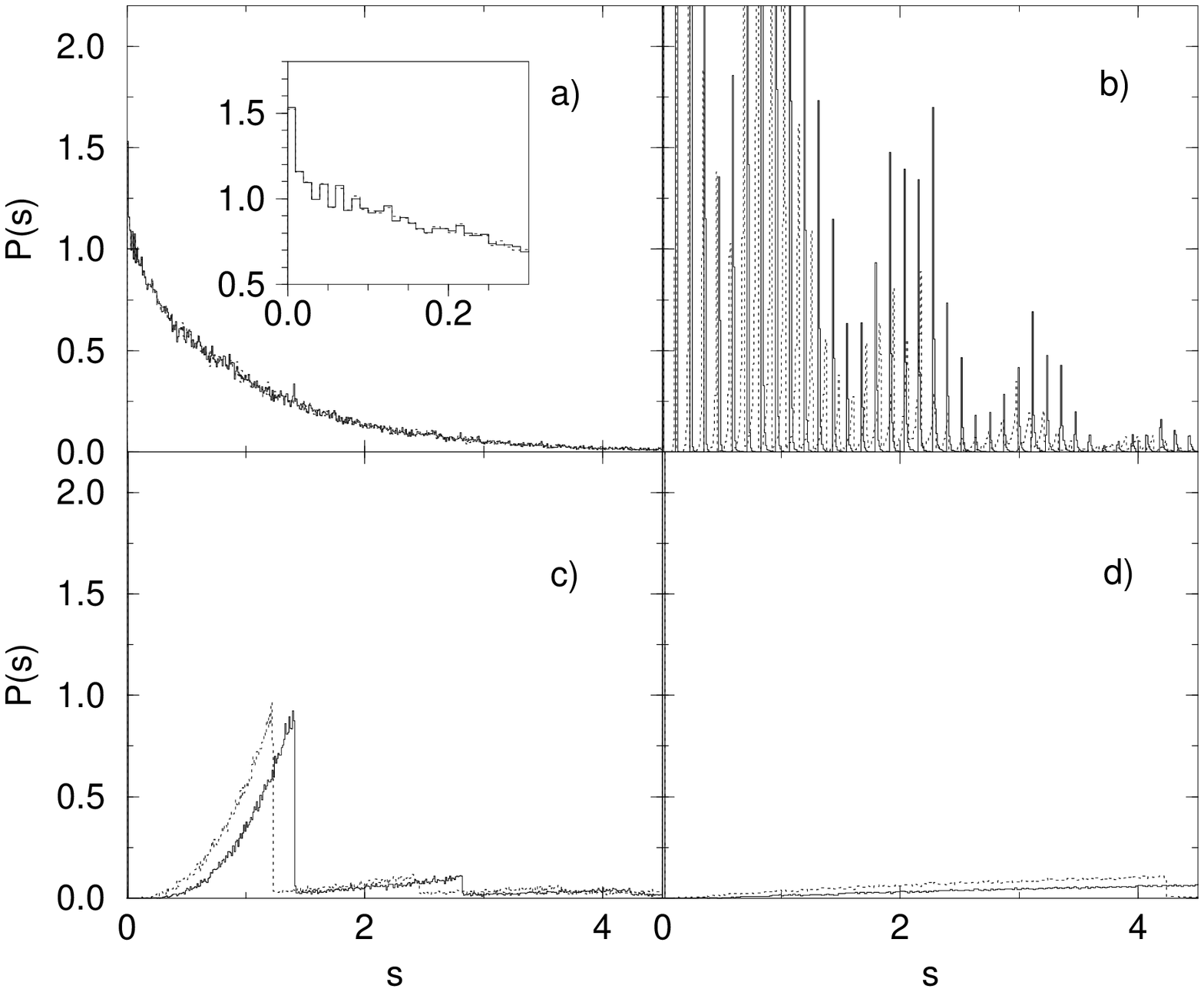}  
\caption[]{Level spacing distribution $P(s)$ for $\alpha=3$ and  
various $D$. The dashed curves refer to the FD, the dotted curves to the  
$BE$ subspaces. (a) $D=2$,  (b) $D=3$, (c) $D=4$ and  (d) $D=5$.  
The number of levels included in each diagram was 
about 96000 for each symmetry subspace. In (c) and (d) there  
is a strong delta function at the origin. The remaining features seen  
are not stationary and disappear as the number of levels is increased.  
\label{ps} }
\end{figure}

As to the case of cubes, the density of two cubes decreases, so that   
the unfolded levels (with mean spacing one) are not simple 
multiples of integers. Thus a continuous range of unfolded 
spacings can be achieved and the level spacing distribution  
is essentially Poissonian, except for a small overshoot at the origin  
(Fig.~\ref{ps}a). \\  
For three cubes, the density is constant, and the spacing   
distribution seems to converge to   
$\delta$-spikes with spacing $1/\rho(E)$. The deviations noticeable in   
Fig.~\ref{ps}b are due to finite $E$ effects: the  
Bose-Einstein spacings have a wing to lower values because of the decreasing  
density, the Fermi-Dirac spacings have one to higher values because of the   
increasing density. However, for four and more cubes, a strong   
$\delta$-function develops at the origin,   
eventually absorbing all the density (Fig.~\ref{ps}c,d).  
  
\subsection{Connection to Warings problem}  
In the case of integer $\alpha$, all eigenenergies are integer. Thus,  
if the density of states increases with energy, eventually one  
will reach a situation where the density exceeds one level per unit   
interval. This critical energy can be calculated to leading order from  
(\ref{density}) to be  
\beq  
E_c(\alpha,D) = \left( {\Gamma\left({\alpha+1 \over \alpha}\right) \over  
D! \Gamma\left({D\over\alpha}\right) } \right)^{-\alpha/(D-\alpha)} \, .  
\label{E_c}   
\eeq  
For $E>E_c$ more and more levels have to fall onto  
the same integer, giving rise to a $\delta$-function at the origin  
in the spacings distribution. This will happen for  
$D>\alpha+1$. The case $D=\alpha$ is marginal. For $\alpha=2$ it is known  
that the spacings between numbers that can be represented as sums of squares  
increases logarithmically, giving rise to a logarithmically increasing  
degeneracy: the spacing distribution converges to a delta function at   
the origin. For $\alpha=3$ and higher, the distribution seems to  
converge to a stick diagram.  
  
Since the density of states increases, one can ask whether all  
integers can in fact be represented as a sum of $D$ integers raised  
to the power $\alpha$. This is Warings problem (Ribenboim 1989).  
More precisely, define  
a number $g(\alpha)$ so that all integers can be represented if  
$D\ge g(\alpha)$. Since some small integers cause special problems,  
Waring considered another number $G(\alpha)$, such that   
if $D\ge G(\alpha)$ then all sufficiently large integers  
can be represented. Obviously $g(\alpha) \ge G(\alpha)$ and   
$G(\alpha)\ge\alpha+1$ because of the above density argument.  
Some results are collected in table~1.  
  
\begin{table}  
\caption[]{Results on Warings numbers, taken from (Ribenboim 1991).  
The probabilistic lower   
estimate is often found to be too optimistic. The critical energies  
$E_c$ (computed using \ref{E_c})  
in the last two columns increase rather rapidly. Some numerical  
consequences of this will be studied in section \ref{pair_corr}.  
\label{Waring}}  
\begin{tabular}{c|c|c|c|c}  
$\alpha$ & $g(\alpha)$ & $G(\alpha)$ & $E_c(D=\alpha+1)$  
& $E_c(D=\alpha+2)$  \cr\hline  
  2 & 4 & 4 \cr  
  3 & 9 & $4\le G\le  7$ & $3.8\cdot 10^4$ & $2.6\cdot 10^3$ \cr  
  4 &19 & $16$           & $1.0\cdot 10^9$ & $1.3\cdot 10^6$\cr  
  5 &37 & $6\le G\le 21$ & $1.6\cdot 10^{15}$ & $6.0\cdot10^9$\cr  
  6 &73 & $9\le G\le 31$ & $2.4\cdot 10^{23}$ & $2.8\cdot 10^{14}$\cr  
  7 &$143\le g \le 3806 $ &$ 8 \le G\le 45$ & $4.6\cdot 10^{33}$  
& $1.6\cdot 10^{20}$ \cr  
  8 &$279\le g \le 36119$ &$32 \le G\le 62$ & $1.4\cdot 10^{46}$   
& $1.3\cdot 10^{27}$ \cr  
  9 &$548\le g $          &$13 \le G\le 82$ & $9.0\cdot 10^{60}$   
& $1.6\cdot10^{35}$\cr  
\end{tabular}   
\end{table}  

The considerations of the permutation symmetry give rise to a  
specialization. Since both the Maxwell-Boltzmann case (which agrees with  
the sums Waring considered) and the Bose-Einstein statistics allow  
for repetition of integers, Warings numbers remain unchanged. However,  
the Fermi-Dirac statistics poses the additional constraint that all  
the integers have to be different. Thus it is not possible  
to fill in 1's if there is still a small gap to an integer. Therefore,  
the equivalent of $g(\alpha)$ makes no sense, some small integers  
will always be missed. However, the density of states still  
increases for $D\ge\alpha+1$ without bound, so that the density  
of points is sufficient to reach all larger integers. Numerical  
tests suggest that at least for $\alpha=3$ and $D=4$ and $5$ the density of   
points not represented among the lowest $10^6$ integers decreases, 
but that is insufficient since the numbers involved rapidly grow  
large. It might be interesting to study the existence and values of  
$\tilde G(\alpha)$, such that for $D\ge\tilde G(\alpha)$ all sufficiently  
large integers can be represented as sums of $\alpha$'s powers of   
$D$ {\em different} integers. Obviously, $\tilde G(\alpha)\ge G(\alpha)$.  
  
\section{Form factor and pair correlations}  
\subsection{Numerical Results}  
The level spacing distribution $P(s)$ is a complicated mixture of  
$n$-point correlation functions, $n=2,3,\ldots$ and thus not accessible   
to a complete semiclassical analysis. Some progress can be made  
for the two point correlation function  
\beq  
C(\epsilon) = \left\langle\rho(E+\epsilon/2) \rho(E-\epsilon/2)\right\rangle  
/ \langle \rho \rangle^2  
\label{Ce}  
\eeq  
and the derived quantities spectral rigidity  and number variance  
(Berry 1985, Seligman and Verbaarschot 1987, Verbaarschot 1987,  
Bohigas 1991). The Fourier transform of the correlation function 
(\ref{Ce}), the form factor, is the absolute value square of 
the Fourier transform of the spectrum. Since the latter can 
be related to periodic obits via the Berry-Tabor (1977b) semiclassical 
expansion or Poisson summation, a combination of classical 
results (Hannay and Ozorio de Almeida 1984) and quantum information 
(Berry 1985) can be used to estimate the form factor. 
                                                       
\subsection{Berry-Tabor expansion and Poisson summation formula}  
The density of states in the Bose-Einstein and Fermi-Dirac subspaces is to   
leading order given by that for the Maxwell-Boltzmann case with corrections  
for energy levels where two or more quantum numbers coincide   
(cf. \ref{dos}). The Berry-Tabor (1977b) semiclassical expansion in terms  
of classical periodic orbits for the full density is thus a superposition  
of the ones for $D$ or fewer particles with Maxwell-Boltzmann statistics.  
Technically, the semiclassical expansion of Berry and Tabor reduces  
to a Poisson summation on the EBK quantized eigenvalues,  
\beq  
\sum_{n_i} \delta \left(E-\sum_{i=1}^D n_i^\alpha\right)  
= \sum_{m_i} g(m_1,\ldots,m_D)  
\eeq  
where  
\beq  
g(m_1,\ldots,m_D) = \int d^D x  \Pi_{i=1}^D \Theta(x_i)  
\delta\left(x-\sum_{i=1}^D n_i^\alpha\right)  
e^{2\pi x\cdot m} \, ,  
\eeq  
and  
\beq  
\Theta(x) = \left\{ \matrix {1 & x>0 \cr 1/2 & x=0 \cr 0 & x<0} \right.    
\eeq 
is the Heaviside step function. The Fourier transform of the product of step  
functions gives rise to another set of corrections to the leading  
order term, which can be combined with the ones due to symmetrization. 
  
Going through the algebra of expressing the delta function by its  
Fourier representation, doing the $x$-integrals   
and then the final $k$-integral in stationary phase, one ends up with   
an expression of the form (Seligman and Verbaarschot 1987) 
\bea  
g(m_1,m_2,\ldots,m_D) &\sim&  
%{1\over \alpha} \left(-{i\over\alpha-1}\right)^{(D-1)/2}  
\left({\Large \Pi} m_i\right)^{-(\alpha-2)/(2(\alpha-1))}  
%\,\cdot\nonumber\\  
L^{(\alpha-1-D)/(2\alpha)}   
\,\cdot\nonumber\\ &\, &  
E^{-(2\alpha-D-1)/(2\alpha)}   
\exp\left({2\pi i L^{(\alpha-1)/\alpha} E^{1/\alpha}}\right)  
\eea  
where  
\beq  
L = \sum m_i^{(\alpha/(\alpha-1)} \,.  
\eeq  
Semiclassically, the $L(m_1,\ldots,m_D)$ are the periods of the   
classical motions. The lower order terms where one or two quantum   
numbers are equal are of similar form, but with coefficients  
multiplying the powers of $m_i$.  
  
The main conclusion to be drawn from this is that the exponents  
$\alpha$ and $\tilde\alpha=\alpha/(\alpha-1)$ are conjugate  
to one another. In particular, $\alpha={p+q\over p}$ corresponds to   
$\tilde\alpha = {p+q\over q}$. Thus the periods for the eigenvalues  
with $\alpha=3/2$ are sums of cubes. Since sums of four or more 
cubes show large degeneracies, the periodic orbit spectrum for this 
system is degenerate which should influence the  
pair correlation function. 
 
\subsection{Pair correlations}  
\label{pair_corr}  
By the Berry-Tabor (1977b) expansion the density of states can   
be written 
\beq  
\rho(E) = \sum_{\bf m} g_{\bf m} e^{2\pi S_m E^{1/\alpha}}  
\eeq 
with the actions 
\beq 
S_m = L(m)^{(\alpha-1)/\alpha} 
\eeq 
and the amplitudes given above.  
The Fourier transform  
in the scaling variable $x=E^{1/\alpha}$ (alternativly one can 
expand around a reference energy $E_0\rightarrow \infty$  
and consider a small interval around $E_0$) then consists of  
$\delta$-functions at actions $S_m$. 
When taking the absolute value square, there are 
two contributions, one from every orbit with itself and 
one from different orbits. If there are no degeneracies 
in orbit actions, the diagonal part gives the constant 
form factor expected for Poissonian distributed levels 
(Berry 1985). However, with degeneracies as in the  
case $\alpha=3/2$, there are 
additional contributions from the cross terms and the form factor 
is higher by a factor $\overline{g}$, the mean degeneracy of 
actions (Biswas et al 1991).  
  
The diagonal approximation can be good at best up to a period $T_H\sim  
\hbar \rho(E)$, for then the individual eigenvalues can be resolved and  
a quantum sum rule predicts the saturation to a constant, the   
Fourier transform of a $\delta$-function (Berry 1985).  
Since the degeneracy increases slowly with the period, its 
effect will only be noticeable if the density of states is sufficiently 
high so that for orbits of period near $T_H$ the degeneracies 
will become important. For the lowest 100000 states studied here 
and in the studies of Biswas et al (1991), no effect of periodic 
orbit degeneracies on pair correlation functions was found. 
This may have to do with the large numbers involved before 
the mean density of states is so high that degeneracies are enforced. 
This problem awaits further study.

\end{document}